\documentclass[epjCONF,columns]{svjour}
\usepackage{graphics}
\usepackage[varg]{txfonts} 
\usepackage[latin1]{inputenc}
\session-title{2011 Hadron Collider Physics symposium (HCP-2011)}
\begin{document}
\title{Search for Standard Model Higgs boson in the
two-photon final state in ATLAS}
\author{Olivier Davignon\thanks{\email{davignon@cern.ch}} on behalf of the ATLAS Collaboration}
\institute{LPNHE des Universit\'es Paris VI et Paris VII, IN2P3/CNRS, F-75252 Paris, France}
\abstract{
We report on the search for the Standard Model Higgs boson decaying into two photons based on proton-proton collision data with a center-of-mass energy of $7$ TeV recorded by the ATLAS experiment at the LHC. The dataset has an integrated luminosity of about $1.08\;fb^{-1}$. The expected cross section exclusion at $95\%$ confidence level varies between $2.0$ and $5.8$ times the Standard Model cross section over the diphoton mass range $110 - 150$ GeV. The maximum deviations from the background-only expectation are consistent with statistical fluctuations.
}

\maketitle
\section{Introduction}
\label{intro}
In the Standard Model (SM), the introduction of a scalar field with nonzero average value is responsible for the spontaneous ElectroWeak Symmetry Breaking (EWSB), giving masses to the intermediate $W^{\pm}$ and $Z^0$ bosons and regularizing the weak bosons quartic coupling divergence at high energies \cite{Higgs}. This scalar field leads to the theoretical existence of a further elementary massive quantum particle called the Higgs boson. In spite of efforts of past experiments at $LEP$ and the $TeVatron$, the Higgs boson has not yet been observed \cite{LEP},\cite{TeVatron}. Several mass ranges have however been excluded at $95\%$ Confidence Level, including Higgs mass regions below $114.4$ GeV and from $141$ to $476$ GeV \cite{CONFHCP},\cite{Rolandi}.

\section{SM Higgs boson decaying to two photons}
\label{sec:1}
Following the Standard Model predictions, the Higgs boson can be produced by the LHC proton-proton collision by several processes. The main ones are: the gluon fusion through a quark loop, the vector boson fusion, the associate production with a $W^{\pm}$ or $Z^0$ and the associate production with a top-antitop pair. The corresponding production cross-sections as a function of mass are documented in Ref. \cite{XS}. The SM Higgs boson can decay to a multitude of final states, including a pair of bottom quarks, tau leptons or intermediate bosons, etc. Despite its small branching fraction of $\mathcal{O}(10^{-3})$, the diphoton final state has the advantage of allowing the full reconstruction of the Higgs candidates mass while keeping the background to an acceptable level.

\section{Higgs candidates selection}
\label{sec:2}The results presented here can be found in Ref. \cite{Past}. The ATLAS detector is used to reconstruct pp collision events given by the LHC. They are selected by a primary diphoton trigger requiring two photon objects with a transverse energy $E_{\mathrm{T}}$ greater than $20$ GeV. Events are required to have at least two offline reconstructed photons. The final dataset used in the diphoton analysis represents a total integrated luminosity of approximately $1.08\;fb^{-1}$.\\

The event selection relies on several ATLAS subdetectors: the pixel detector, the semiconductor tracker, the transition radiation tracker and the electromagnetic calorimeter where the photons deposit most of their energies. Both converted and unconverted photons may enter the dataset. One event must have at least one primary vertex with 3 associated tracks, for which the individual transverse momentum $p_{\mathrm{T}}$ is greater than $0.4$ GeV. The photons, reconstructed from the electromagnetic clusters in the event, are required to be outside of a region $|\eta|\in[1.37,1.52]$ which represents the transition region between the electromagnetic calorimeter's barrel and endcaps, and inside $|\eta|<2.37$. The photon with the highest transverse momentum is called the leading photon, while the one with the second highest is called the subleading photon. The first is required to have a transverse momentum $p_\mathrm{T}^{\mathrm{lead\;\gamma}}$ greater than $40$ GeV while the second $p_\mathrm{T}^{\mathrm{sublead\;\gamma}}$ has to be greater than $25$ GeV. The photons are asked to be isolated. The isolation is computed from the energy contained in a $(\eta,\phi)$ circle around the photon candidate, the core region of the electromagnetic cluster being excluded from the calculation. For each photon, it is required that $E_\mathrm{T}^{\mathrm{iso}}<5$ GeV. Both photons are asked to pass \verb+tight+ selection criteria, which relie on shower shape variables and energy leakage inside the hadronic calorimeter. These identification criteria have an efficiency of about $80\%$ for true photons. Finally, only the events with $100<m_{\gamma\gamma}<160$ GeV enter the final dataset. Events passing this offline selection are selected by the trigger with an efficiency greater than $99\%$.\\

\begin{center}
\begin{figure}
\resizebox{0.95\columnwidth}{!}{%
\includegraphics{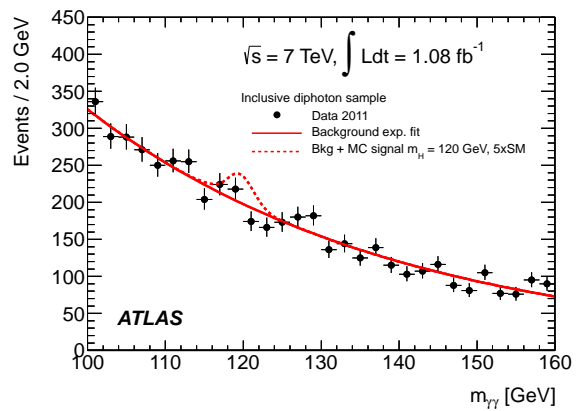}}
\caption{Invariant mass of the diphoton candidates for the 5063 events passing the selection found in the $1.08\;fb^{-1}$ of ATLAS data (black dots). The background model function (defined in Section \ref{sec:4}) is represented by the red line. Finally, five times the SM Higgs signal expectation for $m_H=120$ GeV is represented by the dashed red line.}
\label{fig:4}       
\end{figure}
\end{center}

The final ATLAS dataset consists in $5063$ events, for which the diphoton invariant mass $m_{\gamma\gamma}$ is plotted on Fig. \ref{fig:4}. The background is compared to the dominant $\gamma\gamma$, $\gamma$-jet and jet-jet processes MC predictions. The background level and rejection compare well to the QCD predictions and detector performances \cite{diphoton}. To study the SM Higgs signal, several Monte-Carlo (MC) generators are used. The PowHeg is used to simulate the signal events from the dominant gluon fusion process. MC@NLO is used for cross-checks. PowHeg is also used for the vector boson fusion process, while PYTHIA is used for associate production processes with a vector boson or a top-antitop pair. A full simulation of the detector geometry and response with the GEANT 4 program is performed. The samples are produced so that the multiple interaction pileup is close to the one observed in proton-proton collision data. The same selection is applied on MC and data samples. The generator cross-sections are normalized to the NNLO predictions, except for the $t\bar{t}$ production mode for which only the NLO cross-section is available. The yields of expected SM Higgs signal as a function of the Higgs boson mass are reported in Table \ref{tab:1}. The Higgs signal yield is dominated by gluon fusion while vector boson fusion contributes to about $10\%$ of the total production cross-section.

\section{Data sample composition}
\begin{center}
\begin{figure}
\resizebox{0.95\columnwidth}{!}{%
\includegraphics{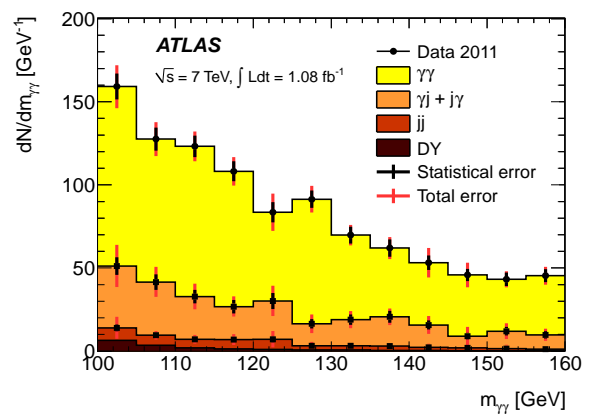}}
\caption{Diphoton, photon-jet, dijet and Drell-Yan contributions to the diphoton candidate invariant mass distribution. The various components are stacked on top of
each other. The error bars correspond to the uncertainties on each component separately.}
\label{fig:8}       
\end{figure}
\end{center}

The main background components to our study are the diphoton production, the photon-jet production for which the jet is misidentified as a photon, the dijet production with two jets misidentified as photons and finally the Drell-Yan continuum where both electrons are misidentidied as photons.\\
To estimate the contributions of the events with at least one jet to the dataset, a method based on control regions of data was used. It relies on two discriminating variables: the isolation and the photon identification cut. The rejection powers of these variables are mostly independent. The method itself is a generalization of the method used in Ref. \cite{2D}. The Drell-Yan background is estimated from $Z\to ee$ candidate events, by drawing the probability of misidentifying one of the two electrons as a photon. This probability is then multiplied to the Drell-Yan predicted yield to evaluate the corresponding background rate.\\
In the diphoton mass range $100$ to $160$ GeV, the number of true diphoton events is found to be  $3650\pm100\pm290$, where the first uncertainty is statistical and the second systematic. The photon-jet and jet-jet contributions are estimated to be $1110\pm60\pm270$ and $220\pm20\pm130$ events, respectively. The Drell-Yan background, which has a softer invariant mass spectrum than the other components, is estimated to $86\pm1\pm14$ events. The different contributions to the diphoton mass range distribution are shown in Fig. \ref{fig:8}. 

\vspace{-30pt}
\section{Event categorization}
\label{sec:3}
Several topological and kinematical event categories are defined in order to benefit from the different signal over background ratios in the different regions of phase space. The present analysis relies on 5 coupled $\eta$ region/conversion status categories:
\begin{enumerate}
\item Unconverted central: both photons unconverted and both located in the central region of the EM calorimeter, i.e. with $|\eta|<0.75$;
\item Unconverted rest: both photons unconverted but at least one is not located in the central region;
\item Converted central: at least one photon is converted but both are located in the central region;
\item Converted transition: at least one photon is converted and at least one photon is located in the region where $1.30<|\eta|<1.75$
\item Converted rest: at least one photon is converted and the event does not lie in the two previous categories.
\end{enumerate}

\begin{table}
\caption{Yields of Higgs signal at the MC generated mass points.}
\label{tab:1}       
\centering
\begin{tabular}{l c c c c c}
\hline
  $m_H$ [GeV] & 110 & 120 & 130 & 140 & 150 \\
  \hline
  Signal yield & 17.0 & 17.6 & 15.8 & 12.1 & 7.7 \\
   \hline
\end{tabular}
\end{table}


\vspace{-25pt}
\section{Background and signal models}
\label{sec:4}
\begin{center}
\begin{figure}
\resizebox{0.95\columnwidth}{!}{%
\includegraphics{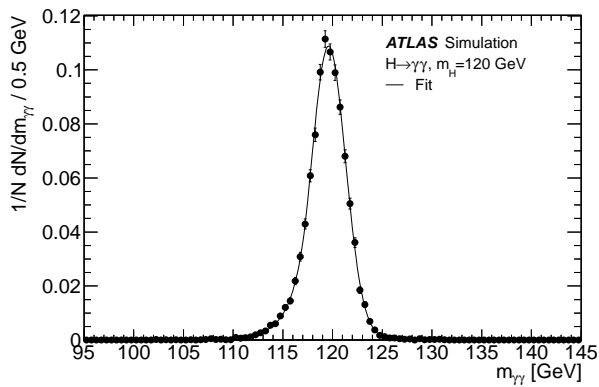}}
\caption{Fit result of the Higgs signal diphoton invariant mass, for $m_H=120$ GeV, for all the categories together. The core component of the mass resolution, $\sigma_{CB}$, is $1.7$ GeV and the FWHM of the distribution is $4.0$ GeV.}
\label{fig:6}       
\end{figure}
\end{center}

The background distributions in each category are estimated by fitting the diphoton mass distribution by an exponential function in the whole selected range.\\
The MC Higgs signal diphoton mass distributions were fitted using a model consisting in a Crystal Ball, to describe the core of the distribution, and a gaussian to describe the tails of the distribution. A fit to the Higgs boson mass shapes is performed on the distributions given by the Monte-Carlo samples. For the simulated point with $m_H=120$ GeV, the core components of the mass resolution have a width which spans between $1.4$ and $2.1$ GeV, depending on the category. The fit result for the $m_H=120$ GeV sample is represented on Fig. \ref{fig:6} for the inclusive sample after selection.

\vspace{-32pt}
\section{Systematic Uncertainties}
\label{sec:5}
As far as Higgs signal is concerned, two types of systematic uncertainties were identified: uncertainties affecting the signal yield and uncertainties concerning the invariant mass resolution. They are summarized in Table \ref{tab:2}. Photon identification and energy resolution systematics dominate. The background modeling uncertainties are estimated by checking how accurately the chosen model fits different predicted diphoton mass distributions and comparing different functional forms for the background model. The resulting uncertainty is between $\pm5$ events at $110$ GeV and $\pm3$ events at $150$ GeV for a Higgs boson mass signal region about $4$ GeV wide.

\begin{table}
\caption{Systematic uncertainties on signal.}
\label{tab:2}       
\centering
\begin{tabular}{l c}
\hline\hline
\multicolumn{2}{l}{\textbf{Uncertainties on the predicted signal yield}} \\
Photon reconstruction and identification (ID) & $\pm11\%$ \\
Isolation cut efficiency & $\pm 3\%$ \\
Trigger efficiency & $\pm 1\%$ \\
Signal cross section & $^{+20}_{-15}\%$ \\
Signal acceptance from modeling of the Higgs $p_\mathrm{T}$ & $\pm 1\%$ \\
Luminosity & $\pm3.7\%$ \\
\hline
\multicolumn{2}{l}{\textbf{Uncertainties on the mass resolution}} \\
Calorimeter energy resolution & $\pm12\%$ \\
Energy calibration extrapolation from $e$ to $\gamma$ & $\pm6\%$ \\
Effect of pileup on energy resolution & $\pm3\%$ \\
Photon angle measurement & $\pm1\%$ \\
\hline
\end{tabular}
\end{table}

\section{Results}
\label{sec:6}
\begin{center}
\begin{figure}
\resizebox{0.95\columnwidth}{!}{%
\includegraphics{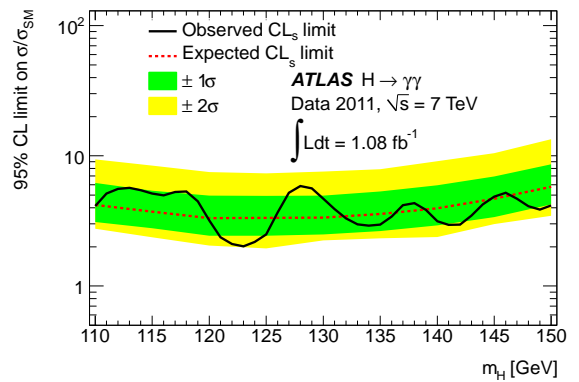}}
\caption{The observed and expected 95\% confidence level limits, normalized to the SM Higgs boson cross sections, as a function of the Higgs boson mass. }
\label{fig:7}       
\end{figure}
\end{center}

The statistical approach used for this analysis is the modified frequentist approach (CLs) documented in Ref. \cite{CLs}. The combined likelihood function is the product of the likelihood functions for the 5 categories. The systematic uncertainties are treated by using nuisance parameters following gaussian PDFs. The $95\%$ confidence level limit on of the inclusive production cross section of a SM-like Higgs boson relative to the SM cross section, is shown on Fig. \ref{fig:7}. The expected CLs spans from $3.3$ to $5.8$ times the SM production cross section for masses between $110$ and $150$ GeV. The observed limits are between $2.0$ and $5.8$ times the SM cross section. No significant excess is observed.

\end{document}